\documentclass[12pt]{article}
\usepackage[centertags]{amsmath}
\usepackage{graphicx}
\newcommand{\be}{\begin{equation}}
\newcommand{\ee}{\end{equation}}
\usepackage[centertags]{amsmath}
\usepackage{amsfonts}

\normalbaselines
\newcommand{\x}{\hat{x}}
\newcommand{\y}{\hat{y}}
\newcommand{\z}{\hat{z}}

\newcommand{\tpartial}{\hat{\partial}}

\begin{document}
\title{\bf Vortex solutions  in the noncommutative torus}
\author{G.S.~Lozano$^a$\thanks{Associated with CONICET}\,,
D.~Marqu\'es$^{b \, *}$ and F.A.~Schaposnik$^b$\thanks{Associated with
CICBA}
\\
{\normalsize\it $^a$Departamento de F\'\i sica, FCEyN, Universidad
de Buenos Aires}\\ {\normalsize\it Pab. I, Ciudad Universitaria,
1428, Buenos Aires, Argentina}
\\
 {\normalsize\it $^b$Departamento de F\'\i
sica, Universidad Nacional de La Plata}\\ {\normalsize\it C.C. 67,
1900 La Plata, Argentina}}

\maketitle
\begin{abstract}
Vortex configurations in the two-dimensional torus are considered
in noncommutative space.  We analyze the BPS equations of the
Abelian Higgs model. Numerical solutions are constructed for the
self-dual and anti-self dual cases by extending an algorithm
originally developed for ordinary commutative space. We work
within the Fock space approach to noncommutative theories and the
Moyal-Weyl connection is used in the final stage to express the
solutions in configuration space.
\end{abstract}
\section{Introduction}

The study of non-trivial classical solutions of field theories defined in noncommutative (NC) space-time has attracted much attention during the last years.
The case of instantons, vortices and monopoles has been analyzed in great detail
~\cite{DN},~\cite{FS}. The analysis of these type of configurations simplifies for particular relations of coupling constants for which it is possible to establish the existence of Bogomolnyi-Prassad-Sommerfeld (BPS) equations \cite{BO}.

For the particular case of vortices in  Abelian Higgs model in the noncommutative plane, the existence of BPS equations was first established in \cite{bps1}-\cite{bps3}.
As in the commutative space counterpart \cite{dVS}, no explicit analytical solutions exist. While in ordinary space the field profiles are found by solving  numerically non linear differential equations \cite{dVS}, in NC space this is done by solving numerically non linear recurrence relations \cite{bps3}.

More recently, BPS equations for the Abelian Higgs Model in a two dimensional torus in NC space were found in \cite{flms}.
As in the case of commuting coordinates, the analysis of solutions to these equations is more complicated than for the planar case due to the presence of non trivial boundary conditions. A very efficient numerical method was recently introduced in ordinary space by Gonzalez-Arroyo and Ramos \cite{GAR}. Guided by this method, we will address in this paper the problem of constructing numerical solutions of the BPS equations in the NC torus.

This paper is organized as follows. In section 2 we introduce gauge
and scalar fields defined in the noncommutative two-torus, discuss
their boundary conditions, and  analyze how gauge covariant and
gauge invariant objects constructed from these fields should be
integrated. The noncommutative Maxwell-Higgs model is introduced in
section 3,
 where the derivation BPS equations and a Bogomolnyi bound for the energy are recalled \cite{flms}. Section 4 is devoted
 to the construction of explicit vortex solutions to the BPS equations, this being achieved
 by using both the Moyal and the Fock space approaches for the treatment of noncommutative systems.

\section{Fields in the noncommutative torus}

We consider noncommutative $2+1$ dimensional space-time with
coordinates satisfying
\begin{equation}
[\x,\y] = i \theta  \label{ncspace}
\end{equation}
\begin{equation}
[\x,t] = [\y,t] = 0 \ ,
\end{equation}
and the space coordinates are defined on a torus, $(\x,\y)\subset {\cal
T}$, the periods of    ${\cal T}$ being $(L_1,L_2)$. We shall be
interested in a $U(1)$ gauge theory with a
 Higgs scalar $\hat{\Phi}$ coupled to
 gauge fields $\hat{A}_i$. The fields transform under the $U(1)$ gauge group
 according to
\begin{align}
\hat{A}_{i} \, &\to \, \hat{A}^{V}_i =  \hat{V}^{-1}\, \hat{A}_{i}
\, \hat{V} + \frac{i}{g}\, \hat{V}^{-1}\,
\tpartial_{i} \, \hat{V} \;\;\;\;\;\;\; i=1,2\\
\hat{\Phi} \, &\to \, \hat{\Phi}^{V} = \hat{V}^{-1}\, \hat{\Phi}\ \ \ \ \ \ \ \ \ \ \ \ \ \ \ \ \ \ \ \ \ \ \ \ \ \ \ \ \ \ \ \ \ \ \ \ \ \ ,
\label{conv1}
\end{align}
with $\hat{V} \in U(1)$ and $g$ the gauge coupling constant.
The fields are functions of ($\x, \y$), that is, they are operators acting on the Fock space generated by Eq.(\ref{ncspace}). As we will
 be looking for static configurations, the time $t$, which in our approach is just a parameter, will not play any role.
For definiteness, we will  consider scalar fields in the fundamental
representation. The other cases (anti-fundamental, adjoint) can be
dealt in a similar way. Here,  derivatives are defined as  in the
noncommutative plane, \be \tpartial_i=\frac{i}{\theta}
\epsilon_{ij}[\x_j,\;\; ]\ . \label{deriv}\ee

As for the ordinary torus, a scalar field on the noncommutative
torus will be defined as a function $\hat{\Phi}(\x,\y)$ which is
periodic up to gauge transformations. That is,
\begin{align}
\hat{\Phi}(\x+L_1,\y) &= \hat{U}_1(\x,\y)\, \hat{\Phi}(\x,\y) = \hat{\Phi}^{(U_1^{-1})}(\x,\y)\nonumber \\%
\hat{\Phi}(\x,\y+L_2) &= \hat{U}_2(\x,\y)\, \hat{\Phi}(\x,\y) =
\hat{\Phi}^{(U_2^{-1})}(\x,\y)\ ,\label{comm-1}
\end{align}
where $\hat{U}_1,\hat{U}_2 \subset U(1)$ are the transition
functions. Accordingly, boundary conditions for gauge fields are
\begin{align}
\hat{A}_{i}(\x+L_1,\y) &=  \hat{A}^{(U_1^{-1})}_i(\x,\y) \\
\hat{A}_{i}(\x,\y + L_2) &= \hat{A}_i^{(U_2^{-1})}(\x,\y)\ .
 \label{bca}
\end{align}
Consistency of the precedent relations leads to an equation for
the $U$'s which is, formally, the same as for the commutative torus,
\be \hat{U}_{2}(\x+L_1,\y)\, \hat{U}_1(\x,\y) =
\hat{U}_1(\x,\y+L_2)\, \hat{U}_{2}(\x,\y)\ . \label{concon}\ee
A solution of this consistency equations is given by
\be \hat{U}_1(\x,\y) = e^{i\, \pi \, \omega\, L_1\, \y} \, ,
\;\;\;\;\;\;\;\; \hat{U}_2(\x,\y) = e^{-i\, \pi \, \omega\, L_2\,
\x}\, , \label{tf} \ee
where
\begin{eqnarray}
 \omega &=& \frac{1}{\theta \pi}  ( 1 - s )  \; , \;\;\;\;\; k \in \mathbb{Z}
 \nonumber\\
 s &=&   \sqrt{ 1 - {2 \pi
\theta k}/{L_1 L_2}}\ \ \ \ \ \ .
\label{tf1}
\end{eqnarray}
In the $\theta \to 0$ limit, Eqs.(\ref{tf})-(\ref{tf1}) go smoothly
to the solution corresponding to  the commutative torus.

Calling  ${\cal A}_\theta$ the space of functions defined on ${\cal
T}$,  a generic  periodic function $\hat{f}(\x,\y)\in {\cal
A}_\theta$ can be written in the form
\be \hat{f}(\x,\y) = \sum_{n_1,n_2} f_{n_1 n_2} \langle n_1,n_2\rangle \ , \ee
where we have introduced
\begin{equation}
\langle n_1,n_2\rangle = \exp \left(2\pi
in_{1}\frac{{\hat x}}{ L_{1}}\right)\exp \left(2\pi
in_{2}\frac{{\hat y}}{L_{2}}\right)\ . \,\,\,\,\,
\end{equation}
An integration  in ${\cal A}_\theta$, which we shall
denote as ${\rm Tr}$, can be formally introduced,
\be I[f] = {\rm Tr}_{\mathcal{ T}} \hat{f}(\x,\y) = f_{00}L_1L_2\ .
\label{integral0} \ee
Gauge invariant local objects are periodic in $\cal{T}$ and
integrals of this kind of objects can be calculated according to this rule.
Nevertheless, gauge {\it covariant} quantities $\hat{f}^c$, satisfy
\begin{align}
\hat{f}^c(\x+L_1,\y) &=  \hat{U}_{1}(\x,\y)\, \hat{f}^c(\x,\y) \, \hat{U}^{-1}_1(\x,\y)  \nonumber\\
\hat{f}^c(\x,\y + L_2) &=  \hat{U}_{2}(\x,\y)\,\hat{f}^c(\x,\y) \, \hat{U}^{-1}_2(\x,\y)\ , %
\label{twisted-ad-bc}
\end{align}
and it is simple to show that quantities such as $\hat{f}^c$ are
periodic in the scaled torus $\mathcal{\tilde{T}}$ with periods \be \tilde{L}_i=s L_i \ .\ee
In this case the functions should be expanded in the basis
\begin{equation}
\langle n_1,n_2\rangle^{*} = \exp \left(2\pi
in_{1}\frac{{\hat x}}{ \tilde{L}_{1}}\right)\exp \left(2\pi
in_{2}\frac{{\hat y}}{\tilde{L}_{2}}\right)\ , \,\,\,\,\,
\end{equation}
and the integral should be understood as
\be I[f^c] = {\rm Tr}_{\tilde{\mathcal{ T}}} \hat{f}^c(\x,\y) =f^c_{00}
\tilde{L}_1 \tilde{L}_2\ . \ee
In theories defined in NC space, the more natural ``local" (i.e, before
integration) variables are covariant quantities (for instance the electromagnetic tensor, the energy density, etc). We will see then that the scaled torus
$\tilde{\mathcal{ T}}$ plays a fundamental role.

Notice that the trace operation satisfies ${\rm Tr} (\hat{f}\hat{g}) = {\rm Tr}
(\hat{g}\hat{f}) $ and reduces to the standard integral on ${\cal
T}$ in the commutative limit. One can see that this definition is crucial for  preserving the
cyclic property of the integral (trace) which in turn is essential
in order to derive the equations of motion. For example, given two
functions $\hat{\Phi}_1(\hat{\vec x})$ and $\hat{\Phi}_2(\hat{\vec
x})$ satisfying boundary conditions
\begin{align}
\hat{\Phi}_i(\x+L_1,\y) &= \hat{U}_1(\x,\y)\, \hat{\Phi}_i(\x,\y) \nonumber \\%
\hat{\Phi}_i(\x,\y+L_2) &= \hat{U}_2(\x,\y)\, \hat{\Phi}_i(\x,\y)\;
, \;\;\;\;\; \;\;\;\;\;\;\;\; i=1,2\ ,\label{phi1-2}
\end{align}
one can see that the product
\be \hat{\Phi}_1^{\dagger}(\hat{\vec x})\, \hat{\Phi}_2(\hat{\vec
x}) \ee
is strictly periodic in the torus $\mathcal{T}$, but the transposed
product
\be \hat{\Phi}_2(\hat{\vec x}) \, \hat{\Phi}_1^{\dagger}(\hat{\vec
x})\ , \ee
is not periodic in $\mathcal{T}$ but in the scaled torus
$\mathcal{\tilde T}$.  Nonetheless, as
 proved in \cite{flms}, the cyclic property of the integral is
still valid provided one integrates the first function in
${\mathcal{T}}$ while the second one in ${\mathcal{\tilde T}}$
\be {\rm Tr}_ {\mathcal{ T}}  \left(
\hat{\Phi}_1^{\dagger}(\hat{\vec x})\, \hat{\Phi}_2(\hat{\vec x})
\right) = {\rm Tr}_ {\mathcal{\tilde T}} \left(
\hat{\Phi}_2(\hat{\vec x})\, \hat{\Phi}_1^{\dagger}(\hat{\vec
x})\right)\ . \label{cyclic-g}  \;\;\;\ee
That is, the cyclic property is preserved with the above definition.
In what follows we shall indistinctly denote the trace operation by
${\rm Tr}$ assuming that the integrand is expanded in its natural
domain of periodicity.

So far, we have identified the space coordinate algebra defined in
Eq.(\ref{ncspace}) with the algebra of  creation-annihilation operators
in a Fock space, and  we have taken fields as operators $\hat{\Phi}$
in such a Fock space. As
in the noncommutative plane, instead of working with fields
depending on noncommuting coordinates $\hat{x}_i$, one can work with
ordinary coordinates $x_{i}$ and introduce a noncommutative $*$
Moyal product

\begin{equation}
\Phi(x)\ast \chi(x)=\Phi(x)\exp\left(\frac{i}{2}\theta ^{\mu \nu
}\overleftarrow{\partial _{\mu }}\overrightarrow{\partial _{\nu
}}\right)\chi(x)  \label{PMoyal}\ . \;\;\;
\end{equation}

The  connection between these two formalisms is found via the Weyl connection,  an isomorphism
that  relates the algebra of functions multiplied
with the noncommutative Moyal product and the algebra of operators in Fock space. For
$(x,y) \in R^2$ the relation reads
\begin{eqnarray}
&& \hat{\Phi}( \x , \y ) =
\int\frac{d^{2}k}{(2\pi)^{2}}\tilde{\Phi}(k_1,k_2)e^{-i(k_1 { \x}
 +k_{2} \y)} \nonumber\\
&&  \hat{\Phi}( \x , \y )\hat{\Psi}(\x , \y)
 = \widehat{\Phi*\Psi }( \x , \y )\ \ \ \ \ \ \ \ \ \ \ \ \ \ \ ,\label{conexionweyl}
\end{eqnarray}
where $\tilde \Phi(k_1,k_2)$ is the Fourier transformed of field $\Phi(x,y)$ defined in ordinary space. This
formula can be easily extended to the torus. Indeed, as we signaled
above, any function $\Phi$ which is periodic in a torus $\mathcal{
T}$ can be Fourier expanded as
\begin{equation}
\Phi(x,y) = \sum_{n_1,n_2} {\tilde \Phi}_{n_1 n_2}
\exp \left(2\pi
in_{1}\frac{{x}}{ L_{1}}\right)\exp \left(2\pi
in_{2}\frac{{ y}}{L_{2}}\right)\ .
\end{equation}
Then, Eq.(\ref{conexionweyl}) valid for $R^2$, is replaced in the torus
$\mathcal{T}$ by
\begin{equation}
\hat{\Phi}(\x, \y)=
\sum_{n_{1},n_{2}} {\tilde\Phi}_{n_{1}n_{2}}  \exp \left(\frac{2 \pi^2 i  n_1 n_2\theta}{L_1 L_2}\right)  \langle
n_{1}, n_{2} \rangle\ . \label{prescrip}
\end{equation}
The connection between integration in both approaches is
\begin{equation}
\int d^{2}x \ \Phi(x ,y ) \to 2\pi \theta\ {\rm Tr} (\hat{\Phi})\ .
\label{ConexionIntTraza}
\end{equation}

The Moyal mapping gives us the possibility to work with commuting coordinates.
The difficulty with this approach is that the resulting expressions (and the equations of motion) are highly non local quantities in the sense that they involve derivatives of arbitrary order. As in the case of $R^2$ we will find more convenient to solve the equations of motion in the Fock space formalism, and use the Moyal correspondence to represent  graphically the final results, by connecting operators with functions defined on configuration space.
\section{The Maxwell-Higgs model}

We are interested in a model with  a $U(1)$ gauge field coupled to
a Higgs scalar defined on the noncommutative torus. Dynamics of
the model is governed by the Lagrangian density
\begin{equation}
\hat{L} =   -\frac{1}{4}\,\hat{F}_{\mu\nu} \hat{F}^{\mu\nu} +
(\hat{D}_\mu \hat{\Phi})^{\dagger}\, (\hat{D}^\mu \hat{\Phi}) -
\lambda \, (\hat{\Phi}^{\dagger} \hat{\Phi}-\Phi_0^2)^2\ .
\end{equation}
We are looking for static solutions to the
equations of motion and hence    we can
look for minima of the energy (per unit length)
 \be E = {\rm Tr}  \,  \left( \frac{1}{4}\,\hat{F}_{ij}
\hat{F}_{ij} + (\hat{D}_i \hat{\Phi})^{\dagger}\, (\hat{D}_i
\hat{\Phi}) + \lambda \, (\hat{\Phi}^{\dagger}
\hat{\Phi}-\Phi_0^2)^2\right)\ . \ee
Here, \be \hat{D}_i \hat{\Phi} = \tpartial_i \hat{\Phi} - i g
\hat{A}_i\, \hat{\Phi} \label{cd} \ee
 is the
covariant derivative, and $\hat{F}_{ij}$ is the electromagnetic
tensor \be \hat{F}_{ij} = \tpartial_i \hat{A}_j - \tpartial_j
\hat{A}_i -ig[\hat{A}_i,\hat{A}_j]\ . \ee
Notice that the covariant derivative used in Eq.(\ref{cd})
corresponds to a Higgs-gauge coupling which can be considered as in
the fundamental representation (other choices are possible).

As in the commutative case, the energy can be rewritten using the
Bogomolnyi trick as~\cite{flms},
\begin{align}
E = & {\rm Tr}   \left( \frac{1}{2}\, |\hat{D}_i \hat{\Phi} - i
\gamma \, \varepsilon_{ij}\, \hat{D}_j\hat{\Phi}|^2 +
\frac{1}{4}\,\left( \hat{F}_{ij} - \gamma\, g\, \varepsilon_{ij}
(\hat{\Phi}\, \hat{\Phi}^{\dagger}  - \,
\Phi_0^2) \right)^2 + \right. \nonumber \\
&\left. \left(\lambda - \frac{g^2}{2} \right) \, \left(
\Phi^{\dagger}\, \Phi - \Phi_0\right)^2 -  \gamma \, \frac{g}{2}\,
\Phi_0^2 \,
\varepsilon_{ij} \, \hat{F}_{ij} + \text{total derivative}\right)\ , \nonumber \\
\label{ener2}
\end{align}
where $\gamma=\pm 1$, and $\varepsilon_{12} = 1$.

Then, choosing coupling constants so that  $\lambda = g^2/2$ (the Bogomolnyi point)
one can establish a BPS bound for the energy
\be E \geq -\gamma \frac{g}{2} \Phi_0^2 {\rm Tr}_ {\mathcal{\tilde
T}} \, \varepsilon_{ij}\hat{F}_{ij} \ ,\ee
where we have indicated that the trace is taken in the scaled torus $\tilde{\tau}$ to stress that $\hat{F}_{ij}$ is a covariant object. The bound is attained when the following BPS  first order
equations hold
\begin{eqnarray}
 \hat{D}_i \hat{\Phi} &=&  i \gamma \, \varepsilon_{ij}\,
 \hat{D}_j\hat{\Phi}
\nonumber\\
 \hat{F}_{ij}~ &=& \gamma\, g\, \varepsilon_{ij} (\hat{\Phi}\, \hat{\Phi}^{\dagger}  -
\, \Phi_0^2)\ .  \label{bpsbps}
\end{eqnarray}
The sign of $\gamma$ should be chosen in such a way that the bound is positive.
For definiteness,  we shall set from here on  $\gamma=-1$.

In order solve these equations, let us start by
observing that boundary conditions (\ref{bca}) together with our
choice of transitions functions (\ref{tf}) imply for the gauge
field the following relations
\begin{align}
\hat{A}_{1}(\x+\tilde{L}_1 ,\y) &=   \hat{A}_{1}(\x,\y) \nonumber\\
\hat{A}_{1}(\x, \y +\tilde{L}_2) &=   \hat{A}_{1}(\x,\y) -
\frac{1}{g}\pi \omega L_2 \nonumber\\
\hat{A}_{2}(\x+\tilde{L}_1 ,\y) &=   \hat{A}_{2}(\x,\y) +
\frac{1}{g}\pi \omega L_1 \nonumber\\
\hat{A}_{2}(\x, \y +\tilde{L}_2) &=   \hat{A}_{2}(\x,\y)\ \ \ \ \ \ \ \ \ \ \ \ \ .%
\end{align}
A solution can be written as
\be \hat{A}_i(\x,\y) = \hat{{\tilde A}}_i(\x,\y) + \hat{a}_i(\x,\y)\ ,
\label{generalA}\ee
where $\hat{\tilde A}_i$ are some periodic functions in the scaled torus
${\mathcal{\tilde T}}$,
and  $\hat{a}_i$  chosen as
\be \hat{a}_i = f\, \varepsilon_{ij} \x^j \ee
with
\be f = \frac{1}{g\theta} \left( 1-\frac{1}{s} \right)
\label{f}\ .\ee
The field strength $\hat{F}_{ij}$ can be written as
\be \hat{F}_{ij}= \frac{1}{s} \hat{\tilde F}_{ij} + f_{ij}\ ,
\label{generalAB} \ee
where
\begin{align}
f_{ij} &= \varepsilon_{ij}\, \frac{2 \pi k}{g}\,\frac{1}{ {\tilde
L}_1 {\tilde L}_2}
\end{align}
and
\begin{align}
\hat{\tilde F}_{ij} & = \tpartial_i \hat{\tilde A}_j - \tpartial_j
\hat{\tilde A}_i - i \, \tilde{g} \, [\hat{\tilde A}_i, \hat{\tilde
A}_j]\ .
\end{align}
Here, we have introduced a scaled charge

\be {\tilde g} = s\, g \ . \ee

Let us now study parameterization (\ref{generalA}) in connection
with   gauge trans\-for\-ma\-tions,
\begin{align} \hat{A}^{V}_i &= \hat{V}^{-1}\left(\hat{\tilde A}_i + \hat{a}_i \right)\hat{V} +
\frac{i}{g}\, \hat{V}^{-1}\, \tpartial_{i} \, \hat{V}  \nonumber \\
&=\hat{V}^{-1}\,\hat{\tilde A}_i \,\hat{V}  +
\,f\,\varepsilon_{ij}\, \hat{V}^{-1}\,\x^j \, \hat{V} +
\frac{i}{g}\, \hat{V}^{-1}\, \tpartial_{i} \, \hat{V}\ .
\,\,\,\,
\end{align}
Using Eq.\eqref{deriv} we can rewrite the middle term as a
derivative term plus $\hat{a}_i$
\begin{align}
\hat{A}^{V}_i &= \hat{V}^{-1}\,\hat{\tilde A}_i \,\hat{V}  - i\,
\theta f\, \hat{V}^{-1}\, \tpartial_{i} \, \hat{V}  + \hat{a}_i +
\frac{i}{g}\, \hat{V}^{-1}\, \tpartial_{i} \, \hat{V}  \nonumber \\
 \\
&=\hat{V}^{-1}\,\hat{\tilde A}_i \,\hat{V}  + \frac{i}{{\tilde g}}\,
\hat{V}^{-1}\, \tpartial_{i} \, \hat{V}  + \hat{a}_i\ .
\end{align}
Thus a gauge transformation on $\hat{A}_i$ is equivalent to a gauge
transformation on $\hat{\tilde A}_i$ (keeping $\hat{a}_i$
untransformed) but using the scaled charge ${\tilde g}$ .

We can summarize these results by stating that, in the analysis of
gauge theories in the torus, one can trade
non-trivial boundary conditions in
the noncommutative torus $\mathcal{T}$ by periodic boundary conditions
and a scaled charge ${\tilde
g}$ in a scaled noncommutative torus $\mathcal{\tilde T}$.

Let us now discuss the boundary condition equations for scalar
fields. A field $\hat{\Phi}(\x, \y)$ satisfying the boundary
conditions \eqref{comm-1} with transition functions given by
Eq.\eqref{tf}, can be decomposed as
\be \hat{\Phi}(\x,\y) = \hat{M}^{-1}(\x,\y) \, \hat{\chi}(\x,\y)\ ,
\label{phi-decomp}\ee
where $\hat{\chi}(\x,\y)$ will be an explicit function fixed so
that it satisfies the same boundary conditions as
$\hat{\Phi}(\x,\y)$, and $\hat{M}^{-1}(\x,\y)$, which has periodic
boundary conditions on the torus $\tilde{\mathcal{T}}$, will be
found by solving the equations of motion.

Thus  $\chi$, must satisfy the conditions
\begin{align}
\hat{\chi}(\x+L_1,\y) &= \hat{U}_1(\x,\y)\, \hat{\chi}(\x,\y) \nonumber \\%
\hat{\chi}(\x,\y+L_2) &= \hat{U}_2(\x,\y)\, \hat{\chi}(\x,\y)\ .
\label{app2-bceta}
\end{align}
Using complex variables
\be
\z=\x+i\y , \,\,\,\,\,\, \bar{\z}=\x-i\y\ ,
\ee
a solution can be written as \cite{flms}
\be \chi(\z,\bar{\z}) = {\mathcal{N}} e^{\frac{\alpha}{2} \left\{\z\, , \,
\z-\bar{\z}\right\} } \, \prod_{n=1}^{|k|} \hat{\theta}_3\left(\pi
(\z + a_n)/L_1 |i L_1/L_2\right)\ . \label{eta-k} \ee
Here $\hat{\theta}_3(\z|\tau)$ is the Riemann $\theta$ function \be
\hat{\theta}_3(\z|\tau) = \sum_n e^{i\pi \tau n^2 + 2 i n \z}\ ,
\label{theta-3}\ee the  $a_i$ are $|k|$ complex constants (which
will be associated with the center of the vortices) satisfying \be
\sum_{n=1}^{|k|} a_n = 0\ , \ee
 and
\be \alpha= -\frac{1}{2 \theta} \log\left(1 - \pi \omega
\theta\right) = -\frac{1}{2 \theta} \log s \ .\ee
In the $\theta \to 0$ limit,  this function coincides with the one
obtained in the commutative case (see \cite{GAR}). Note that in the
particular case of $k=1$, Eq.(\ref{eta-k}) simplifies to
\be \hat{\chi}(\z,\bar{\z}) = {\mathcal{N}} e^{\frac{ \alpha}{2} \left\{\z\, , \,
\z-\bar{\z}\right\} } \, \hat{\theta}_3\left(\pi \z /L_1 |i
L_1/L_2\right)\ . \label{etak1}\ee
%


~

Using decomposition (\ref{generalA}) the BPS equations can be rewritten in the form
\begin{eqnarray} \hat{\tilde F}_{z \bar{z}} &=& \, i \frac{\tilde g}{2} \left(
\left(\Phi_0^2 - \frac{2 \pi k}{ g^2 {\tilde L}_1 {\tilde
L}_2}\right)  -  \hat{\Phi}\hat{\Phi}^\dagger\right) \label{bo1}
\\
 \hat{\tilde D}_{\bar z}\hat{\Phi} &=& - \frac{\pi \, \omega}{2} \hat{\Phi} \, \z\ \ \ \ \ \ \ \ \ \ \ \ \ \ \ \ \ \ \ \ \ \ \ \ \ \ \ \ \ ,
\label{bo2}
\end{eqnarray}
where
\begin{equation}
\hat{\tilde F}_{\z \bar{\z}} = \hat{\partial}_{z}\hat{\tilde{A}}_{\bar{z}} - \hat{\partial}_{\bar{z}}\hat{\tilde{A}}_{z} - i \tilde{g} \left[\hat{\tilde{A}}_{z}, \hat{\tilde{A}}_{\bar z}\right]\ ,
\end{equation}
and the complex gauge fields are defined as $\hat{\tilde{A}}_{z} = \frac{{\hat{\tilde{A}}}_1 - i {\hat{\tilde{A}}}_2}{2}$.
Since the fields $\hat{\tilde A}_i$ are periodic in the scaled torus
$\mathcal{\tilde T}$, the total flux ${\cal F}$ of $\hat{\tilde
F}_{ij}$ on $\mathcal{\tilde T}$ vanishes (see equation
\eqref{integral0}), and then  we have
\begin{align}
{\cal F} &={\rm Tr}_ {\mathcal{\tilde T}} \,  \hat{F}_{12} = {\rm
Tr}_ {\mathcal{\tilde T}} \,  f_{12} =  \frac{2 \pi k}{g}\ .
\end{align}

It is easy to see that the ansatz (\ref{phi-decomp}) automatically
satisfies the BPS equation (\ref{bo2}) provided that the gauge field
is chosen as \be \hat{\tilde A}_{\bar z} = \frac{i}{\tilde g}
\hat{M}^{-1} \tpartial_{\bar{z}} \hat{M}\ , \label{ParA}\ee where
$\hat{M}$ is a  (non unitary) function periodic in $\mathcal{\tilde
T}$.

Then, it only  remains to find $\mathcal{N}$ and $\hat{M}$ appearing in Eqs.(\ref{eta-k})-(\ref{ParA}) so that the gauge and scalar fields
satisfy the BPS equation (\ref{bo1}).
Defining
\be \hat{H}=\hat{M}\, \hat{M}^{\dagger}\ , \ee
the field strength  $\hat{\tilde F}_{z\bar z}$ can be written
as \be \hat{\tilde F}_{\z {\bar \z}}  = \frac{i}{\tilde g} \hat{M}
^{-1}\,\hat{H}\, \tpartial_{z}\left(\hat{H}^{-1}\tpartial_{\bar z}
\hat{H} \right) \hat{M}^{\dagger {-1}}\  \ee
and Eq.(\ref{bo1}) takes the form
\be \hat{H}\, \tpartial_{z}\left(\hat{H}^{-1}\tpartial_{\bar z}
\hat{H} \right)=
 \frac{1}{2} \,{\tilde g}^2 \left(\epsilon \hat{H} -  \hat{\chi}\, \hat{\chi}^{\dagger}\right)\ ,
  \label{chose}
 \ee
where
\be \epsilon = \Phi_0^2 - \frac{2 \pi k}{ g^2 {\tilde L}_1 {\tilde
L}_2}\ . \ee
It is convenient at this point to introduce
dimensionless fields and variables defined as
\be \hat{\Phi} \to \frac{1}{\Phi_0} \hat{\Phi},  \,\,\, \hat{A}_i
\to \frac{1}{\sqrt{2} \Phi_0} \hat{A}_i,  \,\,\, \x_i  \to
\sqrt{2} g {\Phi_0} \x_i\ , \,\,\, \ee
and redefine the parameters \be \lambda \to \frac{2}{g^2} \lambda,
\;\;\; \theta \to 2g^2 \Phi_0^2 \theta \ .\ee
With these conventions, the Bogomonlyi point corresponds to $\lambda = 1$, and Eq.(\ref{chose}) becomes
\be \hat{H}\, \tpartial_{z}\left(\hat{H}^{-1}\tpartial_{\bar z}
\hat{H} \right)=
 \frac{s^2}{4} \left(\epsilon \hat{H} -  \hat{\chi}\, \hat{\chi}^{\dagger}\right)\ ,
\label{eqfund1}
  \ee
where we have redefined $\varepsilon$ as
\begin{equation}
\varepsilon = 1 - \frac{4 \pi k}{\tilde{L}_{1} \tilde{L}_{2}}\ .
\end{equation}

In what follows we shall discuss in detail the numerical method
used to solve this equation.

\section{Constructing solutions}

In order to find  solutions to the BPS equations  we shall first find
$\hat{H}$ satisfying Eq.(\ref{eqfund1}) and  determine $\mathcal{N}$, the
Higgs field normalization constant. We first need to compute, within the operator approach, the Fourier expansion of $\hat{\chi} \hat{\chi}^\dagger =
\mathcal{N}^{2}\hat{\eta}\hat{\eta}^\dagger$,
\begin{equation}
\hat{\eta} \hat{\eta}^\dagger = \sum_{n_{1},n_{2}}
\eta_{n_{1},n_{2}}\langle n_{1},n_{2}\rangle^*\ .  \label{expxsi2}
\end{equation}
Notice that the domain of periodicity is $\mathcal{\tilde{T}}$, then the appropriate basis
is
\be
\langle
n_{1},n_{2}\rangle^* =\exp \left(2\pi
in_{1}\frac{{\hat x}}{s L_{1}}\right)\exp \left(2\pi
in_{2}\frac{{\hat y}}{s L_{2}}\right)\ .
\ee
Surprisingly, it is possible to find as in the commutative space case, a closed expression for this quantity. Indeed,
using the definition of $\hat{\eta}$ (Eq.(\ref{etak1}) for the single-vortex case $k = 1$), and after a long calculation, the coefficients $\eta_{n_{1},n_{2}}$ (properly
normalized as $Tr_{\tilde{\tau}}(\hat{\eta}\hat{\eta}^\dagger) =
\tilde{L}_{1}\tilde{L}_{2}$) can be written as
\begin{eqnarray}
\eta _{n_{1},n_{2}}=(-1)^{n_{1}n_{2}}e^{ -\frac{\pi}{2s^{2}}
\left(\frac{L_{2}^{2}n_{1}^{2}+
L_{1}^{2}n_{2}^{2}}{L_{1}L_{2}}\right)} e^{2 \pi^2 i n_1 n_2
\frac{\theta} {s^2 {L}_1 {L}_2}}\ .\label{CoefEtan1n2}
\end{eqnarray}

To construct solutions to  Eq.(\ref{eqfund1}) we shall extend the
technique described in \cite{GAR} for the commutative torus to the
noncommutative case.  Since for $\varepsilon = 0$ there is a trivial
solution \be \hat{H} = Constant,  \,\,\,\,\, {\cal N} = 0\ , \ee
 we can use $\varepsilon$ as a perturbative parameter and
expand $\hat{H}$ and the normalization constant $\mathcal{N}$  in
powers of $\varepsilon$

\begin{equation}
\hat{H}=\sum_{k=0}^{\infty}\hat{H}_{k}\varepsilon^{k},  \,\,\,\,
\hat{H}^{-1}=\sum_{k=0}^{\infty}\hat{\bar{H}}_{k}\varepsilon^{k},  \,\,\,\,
{\mathcal{N}}^{2}=\sum_{k=0}^{\infty }A_{k}\varepsilon ^{k}\ .  \label{N2eps}
\end{equation}%
Coefficients $\hat{H}_k$ and $\hat{\bar{H}}_k$, are operators
periodic in $\tilde{\mathcal{T}}$ and can then be Fourier expanded,
\begin{equation}
\hat{H}_{k}=\ \sum_{n_{1},n_{2}}h_{n_{1},n_{2}}^{(k)}\langle
n_{1},n_{2}\rangle^*, \,\,\,\,\,
\hat{\bar{H}}_{k}=\ \sum_{n_{1},n_{2}}\bar{h}_{n_{1},n_{2}}^{(k)}\langle
n_{1},n_{2}\rangle^*\ .
 \label{Hka}
\end{equation}
 Inserting these expansions in Eq.(\ref{eqfund1}) one can determine
 order by order the coefficients,
\begin{eqnarray}
h_{n_{1},n_{2}}^{(0)} &=& \bar{h}_{n_{1},n_{2}}^{(0)}=  \left\{
\begin{array}{cc}
1 & n_{1}=\ n_{2}=0 \\ 0 & n_{1}\neq 0,n_{2}\neq 0
\end{array}
\right.
   \label{CoefHCero}
\nonumber\\
\nonumber\\
h_{n_{1},n_{2}}^{(1)}&=& \left\{
\begin{array}{cc}
0 & n_{1}=\ n_{2}=0 \\
\pi s^{2}\frac{\eta _{n_{1},n_{2}}}{\eta _{0,0}}\frac{1}{%
| \xi _{n_{1},n_{2}}| ^{2}} & n_{1}\neq 0,n_{2}\neq 0%
\end{array}%
\right.   \label{CoefHUno}
\nonumber\\
\nonumber\\
\bar{h}_{n_{1},n_{2}}^{(1)} &=& - h_{n_{1},n_{2}}^{(1)}\ \ \ \ \ \ \ \ \ \ \ \ \ \ \ \ \ \ \ \ \ \ \ \ \ \ \ \ \ \ \ \ \ \ \ \ \ \ \ ,
\end{eqnarray}
where
\begin{equation}
\xi _{n_{1},n_{2}}\equiv \pi \sqrt{\tilde{T}}\left(\frac{n_{2}}{\sqrt{\tilde{T}}}  + i  n_{1}
\right)\ , \label{epsDeriv}
\end{equation}
with $\tilde{T} = \tilde{L}_{2}/\tilde{L}_{1}$ the aspect ratio of
the scaled torus. In the same way one can calculate coefficients to
any order $N$ in $\varepsilon$
\begin{equation}
h_{n_{1},n_{2}}^{(N)}=\left \{
\begin{array}{cc}
0 & n_{1}=\ n_{2}=0 \\
\frac{C_{n_{1},n_{2}}^{(A)}-C_{n_{1},n_{2}}^{(B)}-C_{n_{1},n_{2}}^{(C)}}{|
\xi
_{n_{1},n_{2}}| ^{2}} & n_{1}\neq 0,n_{2}\neq 0%
\end{array}%
\right.   \ ,\label{CoefHN}
\end{equation}
with
\begin{eqnarray}
C_{q_{1},q_{2}}^{(A)}\!\!\!&=& \!\!\!\sum_{n_{1},n_{2}}
\sum_{k=1}^{N-1}\bar{h}_{n_{1},n_{2}}^{(k)}h_{q_{1}-n_{1},q_{2}-n_{2}}^{(N-k)}
\xi _{q_{1},q_{2}}\bar{\xi }_{q_{1}-n_{1},q_{2}-n_{2}}\exp
\left(i\frac{4\pi ^{2}\theta
}{\tilde{L}_{1}\tilde{L}_{2}}n_{2}(q_{1}-n_{1})\right) \nonumber
\\
C_{q_{1},q_{2}}^{(B)}\!\!\!&=&\!\!\! \pi s^{2}\sum_{n_{1},n_{2}}
\sum_{k=0}^{N-1}\bar{h}_{n_{1},n_{2}}^{(k)}A_{N-k} \eta
_{q_{1}-n_{1},q_{2}-n_{2}}\exp \left(i\frac{4\pi ^{2}\theta }{\tilde{L}_{1}%
\tilde{L}_{2}}n_{2}(q_{1}-n_{1})\right) \nonumber
\\
C_{q_{1},q_{2}}^{(C)}\!\!\!&=& \!\!\!\sum_{n_{1},n_{2}}
\sum_{k=0}^{N-2}\bar{h}_{n_{1},n_{2}}^{(k)}h_{q_{1}-n_{1},q_{2}-n_{2}}^{(N-k-1)}
\xi _{q_{1},q_{2}}\bar{\xi }_{q_{1}-n_{1},q_{2}-n_{2}}\exp
\left(i\frac{4\pi ^{2}\theta
}{\tilde{L}_{1}\tilde{L}_{2}}n_{2}(q_{1}-n_{1})\right) \ .\nonumber\label{CB}
\end{eqnarray}
Coefficients ${\bar h}_{n_{1},n_{2}}$, appearing in the expansion of
$H^{-1}$, are obtained from  $\hat{H}\hat{H}^{-1} = 1$,
\begin{equation}
\bar{h}^{(N)}_{q_{1}, q_{2}} = - \sum_{n_{1}, n_{2}} \sum_{k =
1}^{N} h^{(k)}_{n_{1},n_{2}}\bar{h}^{(N-k)}_{q_{1} - n_{1}, q_{2} -
n_{2}}\exp{\left(4 \pi^{2} i n_{2} (q_{1} - n_{1}) \theta /
\tilde{L}_{1}\tilde{L}_{2}\right)}\ .
\end{equation}

One also has  to find a recurrence relation for the coefficients
$A_N$ appearing in the expansion of  the Higgs field normalization
(Eq.(\ref{N2eps})). For this, one uses the condition
$Tr_{\tilde{\tau}}(\hat{\tilde{F}}_{12}) = 0$ finding
\begin{eqnarray}
A_0 &=& 0 \nonumber\\
A_1 &=& \frac{1}{\eta_{0,0}} \nonumber\\
A_N &=& - \frac{1}{\eta _{0,0}}\sum_{n_{1},n_{2}} \sum_{k=1}^{N-1}
\eta_{-n_{1},-n_{2}} \bar{h}_{n1,n2}^{(k)}A_{N-k}  \exp
\left(-\frac{i4\pi ^{2}\theta
}{\tilde{L}_{1}\tilde{L}_{2}}n_{1}n_{2}\right)\ , \;\;\; N>1 \ .\nonumber\\
\label{AA}
\end{eqnarray}

All these equations can be solved recursively. Assuming $\hat{M}$
Hermitian, it can be expanded in powers of $\varepsilon$ as well as
in Fourier modes, and it is possible from $\hat{H} = \hat{M}^2$ to
find recurrence relations for its coefficients. Finally, once the
Fourier coefficients of $\hat{H}$, $\hat{H}^{-1}$, $\hat{M}$,
$\hat{M}^{-1}$ are known, together with  the normalization constant
$\mathcal{N}$, we can use decompositions \eqref{phi-decomp} and
\eqref{ParA} to calculate all the fields in Fock space. The Weyl
connection for periodic functions \eqref{prescrip} can then be used
to establish the correspondence between operators $\hat{O}(\x, \y)$
and their associated functions $O(x, y)$.

Using Eq.(\ref{bo1}) for $\gamma=-1$ and the fact that
$Tr_{\tilde{\mathcal{T}}}\hat{\tilde{F}}_{12}=0$, it follows that
\be Tr_{\tilde{\mathcal{T}}} \left(\Phi_0^2 - \frac{2 \pi k}{ g^2
{\tilde L}_1 {\tilde L}_2}\right)  =
Tr_{\tilde{\mathcal{T}}}\hat{\Phi}\hat{\Phi}^\dagger \geq 0 \ .\ee
Then \be \Phi_0^2 {\tilde L}_1 {\tilde L}_2 \geq \frac{2\pi k}{g^2}\
, \ee or in terms of the dimensionless variables, \be A \geq 4 \pi k
\left( 1+\frac{\theta}{2}\right) \equiv A_{c}\ . \ee Then, the area
$A$ of the torus $\mathcal{T}$ has to be larger than the critical
value $A_c$ in order for solutions to exist.

\begin{figure}
\centering
\includegraphics[width=15cm]{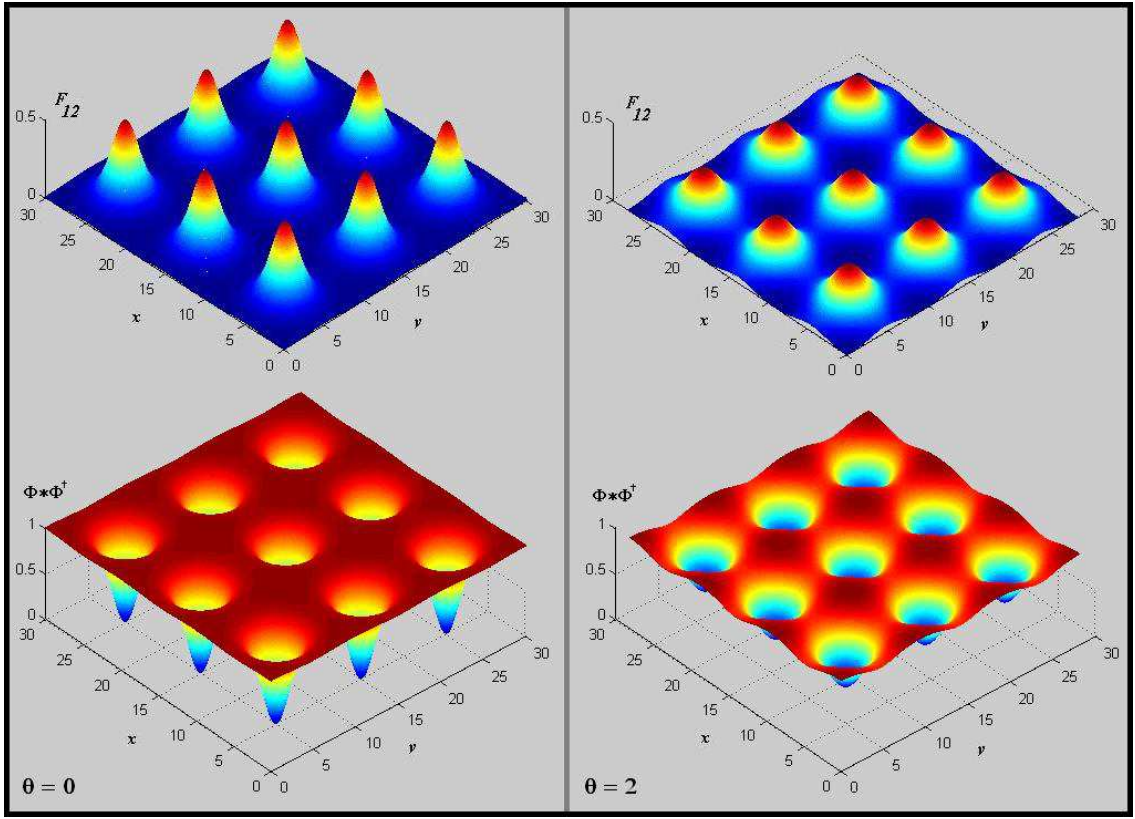}
\caption{\small We represent $F_{12}$ and $\Phi\ast\Phi^\dagger$ for
$\theta = 0$ and $\theta = 2$, for a torus of area $A = 100$. We
consider $3 \times 3$ unitary cells, this leading to an array of $9$
vortices. The distance among vortices equals $\tilde L$, which
explicitly depends on $\theta$. The magnetic flux is always $2\pi
k$, so as in the commutative plane, when incrementing $\theta$, the
vortices change conserving this quantity.} \label{F12T2Lat}
\end{figure}

We will focus first on the single-vortex case $k = 1$, and then
make a few comments on $k> 1$. For simplicity we will only
consider squared torus ($L_1=L_2$). We show in
Fig.(\ref{F12T2Lat}) the solution for $\theta=2$ and $A=100$.
Being the solutions periodic in $\tilde{\mathcal{T}}$, we have
represented them as a lattice of 9 plaquettes being the vortex
solution centered in each plaquette. We plot both quantities
$F_{12}$ and $\Phi\ast\Phi^{\dag}$ (the functions associated
through the Weyl connection to the operators $\hat{F}_{12}$ and
$\hat{\Phi}\hat{\Phi}^{\dag}$ in Fock space). For comparison, we
also show the solutions for the same area and $\theta=0$. In both
cases, $F_{12}$ has a maximum at the center of the torus (the
location of each vortex). Unlike the commutative case,
$\Phi\Phi^{\dag}$ is different form zero at that point.

\begin{figure}
\centering
\includegraphics[width=15cm]{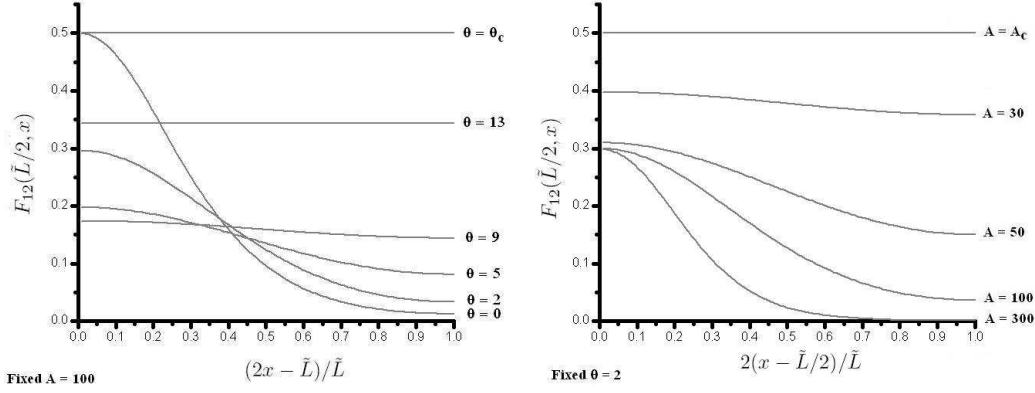}
\caption{\small We show  $F_{12}$ as a function of $x$ (at $y =
\tilde L/2$). The critical values $\theta_{c} \simeq 14$ and $A_{c}
= 8\pi$ correspond to the ($\varepsilon \to 0$) trivial solutions.}
\label{F12100}
\end{figure}

One can study the dependence of the solutions with the area for a
fixed $\theta$. We show in Fig.(\ref{F12100}), $F_{12}$ for several
areas $A \geq A_c=4\pi k(1+\frac{\theta}{2})$ for a fixed
$\theta=2$. Notice that for $A=300$ the configuration is already
similar to the result in the noncommutative plane (see
Ref.\cite{bps3}). Alternatively, we can fix the area $A$ and study
the behavior of the solutions for different $\theta \leq
\theta_c=\frac{A-4\pi k}{2 \pi k}$. We show in Fig.(\ref{F12100})
the results for $A=100$ and several values of $\theta$ in the range
$0\leq \theta \leq \theta_c \sim 14$.

\begin{figure}
\centering
\includegraphics[width=15cm]{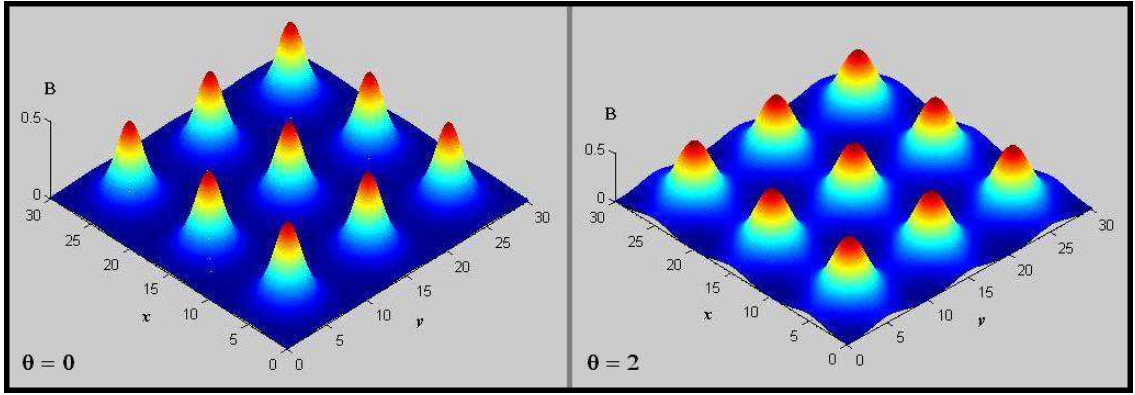}
\caption{\small We show  $\cal
B$ configurations defined on a torus with area $A = 100$ for $\theta=0$ (left) and $\theta=2$ (right). The main difference with $F_{12}$ is
that it is defined on $\mathcal{T}$, and its magnetic flux depends on
$\theta$.} \label{BT2Lat}
\end{figure}

\begin{figure}
\centering
\includegraphics[width=8cm]{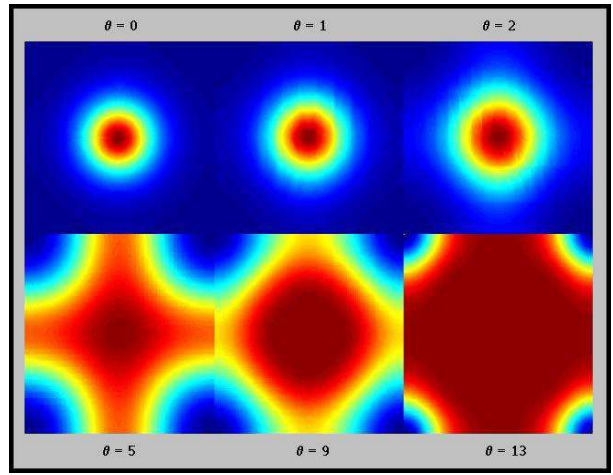}
\caption{\small Upper views of $\cal B$ on a torus with area $A =
100$, for different values of $\theta$. When $\theta \to \theta_{c}
\simeq 14$, ${\cal{B}} \to \frac{1}{2}$.} \label{DeslocSup}
\end{figure}

It is also possible to construct gauge invariant quantities
related to the scalar and gauge fields. Consider for example
\begin{equation}
\hat{\Phi}^{\dag}\hat{\Phi} =
{\mathcal{N}}^{2}\hat{\eta}^{\dag}\hat{H}^{-1}\hat{\eta}\ ,
\end{equation}
and the invariant magnetic field $\hat{\cal{B}}$ (to be distinguished form the covariant one $\hat{F}_{12}$)
\begin{equation}
\hat{\cal B} \equiv
(\hat{\Phi}^{\dag}\hat{\Phi})^{-1}(\hat{\Phi}^{\dag}\hat{F}_{12}\hat{\Phi})
=
(\hat{\Phi}^{\dag}\hat{F}_{12}\hat{\Phi})(\hat{\Phi}^{\dag}\hat{\Phi})^{-1}=
\frac{1}{2}(1 - \hat{\Phi}^{\dag}\hat{\Phi})\ .
\end{equation}
In the $\theta \to 0$ limit, both quantities reduce to their
analogues in commutative space. The flux of $\hat{\cal
B}$ across the torus now depends on $\theta$
\begin{equation}
Tr_{\tau}(\hat{\cal B}) = 2\pi \left(1 + \frac{\theta}{2}\right) k\ .
\end{equation}
We show in Fig.(\ref{BT2Lat}), $\cal B$ configurations in the same conditions of Fig.(\ref{F12T2Lat}).
As it is gauge invariant, it is defined on the same torus
$\mathcal{T}$ for all values of $\theta$ (see Fig.(\ref{DeslocSup})).

We can also consider negative values for the noncommutative
parameter $\theta$. As  the equations remain unchanged when the
following quantities are redefined as \be \gamma \to -\gamma
\nonumber \,\,\, , x_{2} \to - x_{2} \nonumber \,\,\, , A_{2} \to -
A_{2} \nonumber \,\,\, , \theta \to -\theta \nonumber \,\,\, , k \to
-k\ , \label{camvar} \ee this is equivalent to study solutions of the
{\it anti-self dual} equations (this is, BPS Eqs.(\ref{bo1})-(\ref{bo2}) with $\gamma = 1$) but with positive $\theta$ parameter.

In the noncommutative plane, it has been shown that there exists a critical value  $\theta_*$ such that for $\theta<\theta_*$ no solutions to the self-dual equations exist \cite{bps2}, \cite{lmrs}. In the units used in this paper, in the planer case this corresponds to $\theta_*=-2$. The question that arises is if such $\theta_*$ exists also in the NC torus and if it depends on the area.

We have analyzed this problem numerically and we could not make
the method to converge for $\theta < -2$, irrespectively of the value of the area.
 This is completely analogous to what happens in the noncommutative plane
\cite{bps2}, \cite{bps3}, \cite{tong}, \cite{lmrs}, indicating that
$\theta_*=-2$ also for the torus. Thus, we have been able to find
solutions of the self-dual equations $\theta_*<\theta < \theta_c$.
Incidentally, notice that for $\theta=\theta_*$, the critical area
is zero. The case of anti-self dual equations can be considered
using the transformation mentioned above.

\begin{figure}[t]
\centering
\includegraphics[width=15cm]{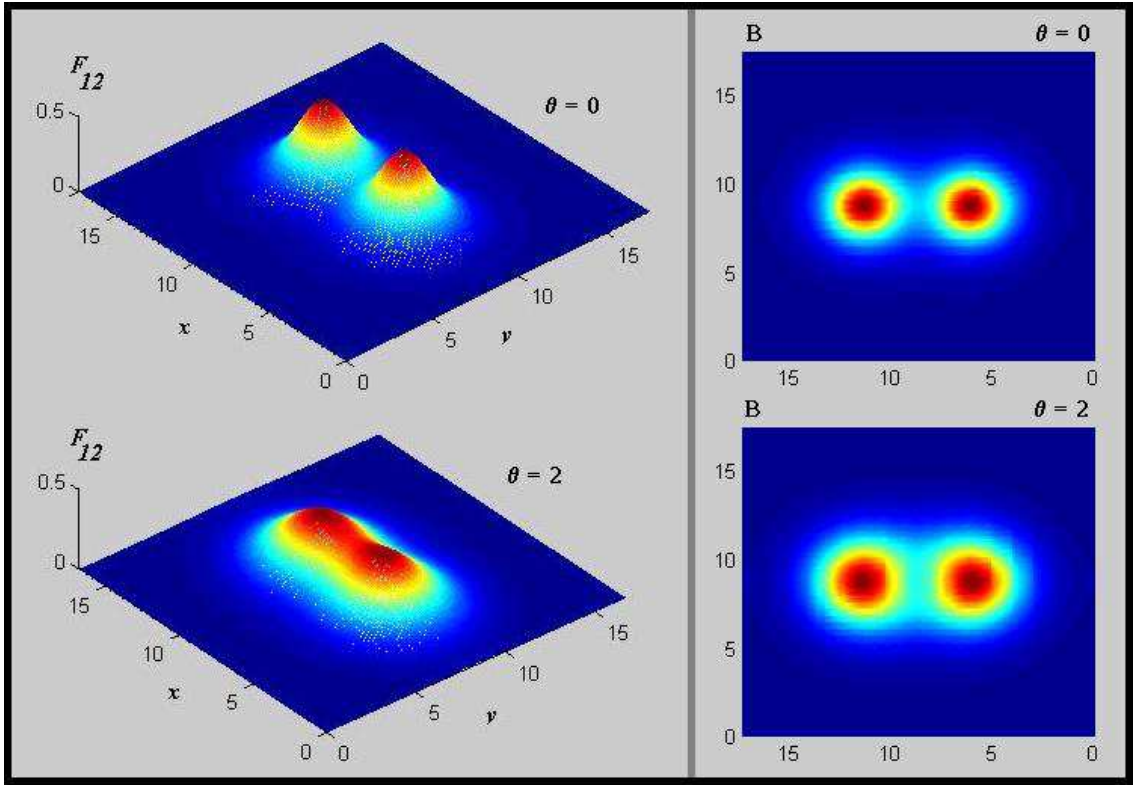}
\caption{\small In the left we represent a two-vortex configuration
of $F_{12}$ for a torus with area $A = 300$, both for $\theta = 0$
and $\theta = 2$. The distance between vortices was fixed to $0.3
L$. In the right we show upper views of $\cal B$ in the same
conditions.} \label{A300T2k2F12}
\end{figure}

It is also possible to obtain solutions for $k = 2$ and higher. In
such cases we do not have closed expressions for the Fourier
coefficients of $\hat{\eta}\hat{\eta}^\dagger$, but numerical
calculations are straightforward. As an example, we show in Fig.(\ref{A300T2k2F12}) $F_{12}$ and upper views of ${\cal B}$ for both $\theta = 0$
and $\theta = 2$.

Finally, let us say a few words about the efficiency of the method.
In order to analyze the accuracy of the algorithm, we have rewritten
Eq.(\ref{eqfund1}), as
\begin{equation}
\frac{1}{\varepsilon}\left[H^{-1}{\cal N}^{2}\eta\eta^{\dag} +
\frac{4}{s^{2}}\partial_{z}\left[H^{-1}\partial_{\bar{z}}H\right]\right]
= 1\ ,
\end{equation}
and verified that the fourier coefficients $A_{q_{1}, q_{2}}$ of the
LHS of this expression
\begin{equation}
A_{q_{1}, q_{2}} = \frac{1}{\varepsilon} \sum_{n_{1}, n_{2}}e^{i4
\pi n_{2}(q_{1} -
n_{1})\frac{\theta}{\tilde{L}_{1}\tilde{L}_{2}}}\left[{\cal N}^{2}\bar{h}_{n_{1},
n_{2}}h_{q_{1} - n_{1}, q_{2} - n_{2}} + \frac{4}{s^2}\bar{h}_{q_{1}
- n_{1}, q_{2} - n_{2}}h_{n_{1}, n_{2}}\bar{\xi}_{q_{1}-n_{1},
q_{2}-n_{2}}\xi_{q_{1},q_{2}}\right]\ ,
\end{equation}
satisfy
\begin{equation}
\sum_{q_1, q_2}\mid A_{q_{1}, q_{2}} - \delta_{q_{1},q_{2}}\mid < 10^{-p}
\end{equation}
for a given p. This is attained by  increasing the number of Fourier coefficients and the order of the
perturbative expansion in $\epsilon$. The same was done for the
$\hat{H}\hat{H}^{-1}=1$ constrain, and for other relevant equations.

Convergence is slower for large area $A \sim 20 A_{c}$ because, as
solutions are more lo\-ca\-li\-zed, more Fourier coefficients are
needed, and besides, $\varepsilon \simeq 1$. For instance, in this
case a convergence for $p = 3$ is achieved with $441$ Fourier modes,
and $N = 300$ orders of the expansion (this demands about $72$ hours
in a standard PC). As the area is reduced, the time of computation
is considerably lower, being of about $20$ seconds for $\varepsilon
\simeq 0$.

As an independent test, we have also calculated the magnetic flux
and energy using the solutions and compared the results with the
analytical values finding an agreement better than ($\sim 10^{-4}$).

 \vspace{1 cm}

\noindent\underline{Acknowledgements}: We wish to thank A.
Gonz\'alez-Arroyo and E. F. Moreno for helpful comments. This work
was partially supported by UNLP, UBA, CICBA, CONICET and ANPCYT.
\newpage




\begin{thebibliography}{99}
%
\bibitem{DN}
  M.~R.~Douglas and N.~A.~Nekrasov,
  Rev.\ Mod.\ Phys.\  {\bf 73} (2001) 977.
  \bibitem{FS}
  F.~A.~Schaposnik,
  Proceedings of the   Second Conference on Fundamental Interactions, Pedra Azul,
Brasil, 2004, Eds. M.~C.~Abdalla et al,
  arXiv:hep-th/0408132.
  %
%
\bibitem{BO} E.~B.~Bogomol'nyi, Sov.\ Jour.\ Nucl.\ Phys. {\bf24} (1976) 1100.
\bibitem{bps1}
D.~P.~Jatkar, G.~Mandal and S.~R.~Wadia,
JHEP {\bf 0009} (2000) 018 ;
\bibitem{bps2}
D.~Bak,
Phys.\ Lett.\ B {\bf 495} (2000) 251 ;
D.~Bak, K.~M.~Lee and J.~H.~Park,
Phys.\ Rev.\ D {\bf 63} (2001) 125010;
\bibitem{bps3}
 G.~S.~Lozano, E.~F.~Moreno
and F.~A.~Schaposnik,
Phys.\ Lett.\ B {\bf 504} (2001) 117;
\bibitem{dVS}
H.~de~Vega and F.A.Schaposnik, Phys.\ Rev.\ D {\bf D14} (1976)
\bibitem{flms}
P.~Forgacs, G.~S.~Lozano, E.~F.~Moreno, F.~A.~Schaposnik,
JHEP 0507 (2005) 074.
\bibitem{tong} D.Tong, J.\ Math.\ Phys. {\bf 44} (2003) 3509-3516
\bibitem{lmrs}G.~S.~Lozano, E.~F.~Moreno, M.~J.~Rodr\'\i guez and F.~A.~Schaposnik,
  JHEP {\bf 0311} (2003) 049
  \bibitem{GAR}A.~Gonzalez-Arroyo and A.~Ramos,
  JHEP {\bf 0407}, 008 (2004).

\end{thebibliography}
\end{document}